# Optical imaging of metallic and semiconductor nanostructures at sub–wavelength regime

A. K. Sivadasan[1], Kishore K. Madapu[1] and Prajit Dhara[2]

[1]Nanomaterials Characterization and Sensors Section, Surface and Nanoscience Division,
Indira Gandhi Centre for Atomic Research, Homi Bhabha National Institute,
Kalpakkam-603102, India

Email: *sivankondazhy@gmail.com*

[2]1st Year Student, Dept. of Electrical and Electronics Engineering
Birla Institute of Technology and Science-Pilani, Pilani Campus
Pilani- 333031,India

**Abstract**

The near–field scanning optical microscopy (NSOM) is not only a tool for imaging of objects in the sub–wavelength limit but also a prominent characteristic tool for understanding the intrinsic properties of the nanostructures. The effect of strong localized surface plasmon resonance absorption of excitation laser in the NSOM images for Au nanoparticles is observed. The role of electronic transitions from different native defect related energy states of AlGaN are also discussed in understanding the NSOM images for the semiconductor nanowire.

## 1. Introduction

The study of light–matter interaction in the near–field regime at the vicinity of nanostructures is a very interesting as well as challenging task for the scientific community. Abbe's diffraction limit prevents conventional optical microscopes to possess a spatial resolution beyond the value of ~$\lambda/2$ (sub–wavelength limit), where $\lambda$ is the wavelength of excitation light with a maximum numerical aperture value of unity for the probing objective. Thus, even for the visible light of $\lambda=400$ nm cannot image nanostructures of size below 200 nm. The near–field scanning optical microscopy (NSOM) assisted with the help of plasmonics is a unique tool to understand the light–matter interaction in the near field regime for optical imaging of nanostructures in the sub–wavelength limit. The light passing through the metal coated tip of NSOM probe with a circular aperture of diameter around few nanometers at the apex is capable of surpassing the diffraction limit [1]. In the near–field regime, the evanescent field emitting from the NSOM probe is not diffraction limited. Hence, it facilitates optical and spectroscopic imaging of objects with nanometer level spatial resolution. Light–matter interactions in metallic nanostructures have opened to a new branch of surface plasmon (SP) based photonics, known as plasmonics. The SPs are originated due to the collective oscillation of the free electrons about the fixed positive charge centers in the surface of metal nanostructures with a frequency of the oscillation of electrons, also known as plasma frequency, $\omega_p = (n_e e^2/m_{eff}\varepsilon_0)^{1/2}$, where, $n_e$ is the density, $m_{eff}$ is the effective mass, $e$ is the charge of an electron and $\varepsilon_0$ is the permittivity of free space [1-3]. The coupling of the incident electromagnetic waves to the coherent oscillation of free–electron plasma near the metal surface is known as a surface plasmon polariton (SPP) and it is a propagating surface wave at the continuous metal–dielectric interface. The electromagnetic field perpendicular to the metal surface decays exponentially and is known as evanescent wave providing sub–wavelength confinement near to the metal surface. Matching of the incident excitation frequency ($\omega$) of electromagnetic wave with the plasmon frequency ($\omega_p$) of the electrons in metal nanostructures, leads to an enhanced and spatially localized light–matter interaction, known as surface plasmon resonance (SPR) [1-3].

AlGaN is an intrinsically $n$–type semiconductor and one of the most prominent candidates among the group III nitride community with wide, direct and tunable band gap from 3.4 to 6.2 eV. Therefore, the group III nitrides including the ternary alloy of AlGaN nanostructures find tremendous applications in short wavelength and high frequency optoelectronic devices including light emitting diodes, displays and optical communications [4]. Consequently, by considering the above mentioned importance of AlGaN nanostructures in optoelectronic applications as well as semiconductor industries, it is also very important to understand the interaction of AlGaN nanowire (NW) with visible light.

In the present report, we have investigated the light–matter interaction of metallic Au nanoparticle (NP) catalysts (diameter ~50–150 nm) along with semiconductor AlGaN NW (diameter ~120 nm) grown *via* vapor liquid solid (VLS) mechanism in the near–field regime by using NSOM technique with external laser excitation of 532 nm (2.33 eV). The variations in contrast and absorption phenomena observed in the NSOM images of Au NPs are understood by considering the plasmonic effects of metallic nanostructures. In order to understand the light–matter interaction of AlGaN NW, we invoked the different energy states related to native defects originating due to the unavoidable incorporation of C and O in the material during the growth process.

## 2. Experimental section

The semiconductor AlGaN NWs along with Au NPs were synthesized using chemical vapor deposition technique *via* VLS growth mechanism. The detailed synthesis and basic characterizations of the sample is available in one of our earlier reports [5].

### 2.1 Atomic force microscopy

The atomic force microscopy (AFM) is one of the types of scanning probe microscopic (SPM) system used for the study of topography related information of a sample with an order of atomic scale spatial resolution. The AFM probes consist with a sharp tip of the order of 100 Å used for probing the tip–sample interactions [6]. There are several interactions possible to contribute the deflection/natural frequency of an AFM cantilever. The common force associated with AFM interaction is inter atomic force called as the van der Waals force and it varies with distance between the tip and the sample (FIG. 1(a)). The three modes of operations, available in the AFM setup (contact, non–contact and intermittent), can be selected by choosing the three different regimes of forces between tip and sample [6]. The major components commonly involved in the SPM systems are shown (schematic, FIG. 1(b)). The SPM system used in the present study works based on tuning fork feedback mechanism. The change in natural frequency of the tuning fork with respect to the tip/sample interactions are considered as a feedback parameter to measure the tip/sample force to map the surface modulation or topography. Thus, the frequency of the tuning fork changes, as the surface morphology changes with the variation of force felt, providing an image of the surface.

The AFM used for the studies on AlGaN NWs, reported in the present studies, is from a SPM system (Nanonics, MultiView 4000; Multiprobe imaging system). The MultiView 4000 uses normal force tuning fork technology with a high Q–factor and phase feedback to allow the control of probe/sample separation. Tuning forks in normal force mode with phase or amplitude feedback permit high performance and ease of operation for AFM imaging in intermittent mode.

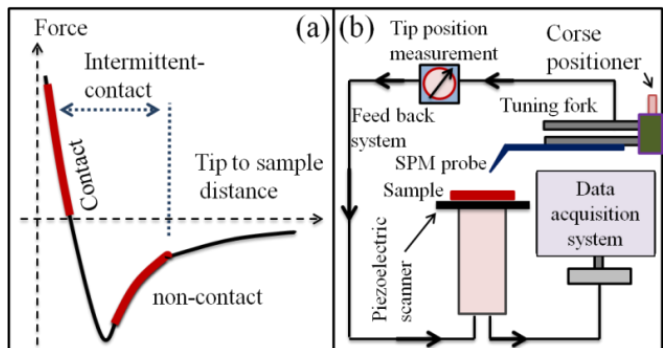

FIG. 1: (a) Interaction force Vs distance between tip and sample. (b) Block diagram of general SPM system.

The AFM tips are glass–based probes. Unlike standard piezo scanners that keep probes separated, the 3D flat scanner with excellent AFM resolution, large vertical (axial) displacement of up to 100 μm for sample scanning and up to 30 μm for tip scanning, is used in this system. The atomic force microscopic (AFM) images are recorded by using a 20 nm tip configured with a tuning fork feedback mechanism (MultiView 4000; Nanonics, Israel).

## 2.2 Near–field scanning optical microscopy

The NSOM is a microscopic technique used to investigate the light–matter interaction of nanostructures in the sub diffraction regime by using the advantages of evanescent waves (or confined light) which surpass the conventional far–field resolution limit. The generation of evanescent waves can be achieved with the help of either by plasmonics or by the use of sub–wavelength apertures coated with noble metals such as Au, Ag or combination of both. The evanescent waves emanated from the apertures/probe with higher momenta *i.e.*, lower wave lengths and velocities compared to that of normal light can be used for achieving the high resolution by placing the detector very close (near–field regime, smaller than wavelength λ) to the sample specimen surface. This allows us to record the light–matter interactions with high spatial, spectral and temporal resolution [1-3]. In this technique, the resolution of the image is determined by the size and geometry of the aperture probe and not by the λ of the excitation light. The NSOM provides simultaneous measurements of the topography and optoelectronic properties of nanostructures with high spatial resolution in the sub–wavelength regime. The detailed schematic of experimental set up used for recording the light–matter interaction of AlGaN single NW is shown (FIG. 2). The NSOM imaging of nanostructures was used to understand the interaction with 532 nm laser (~ 2.33 eV).

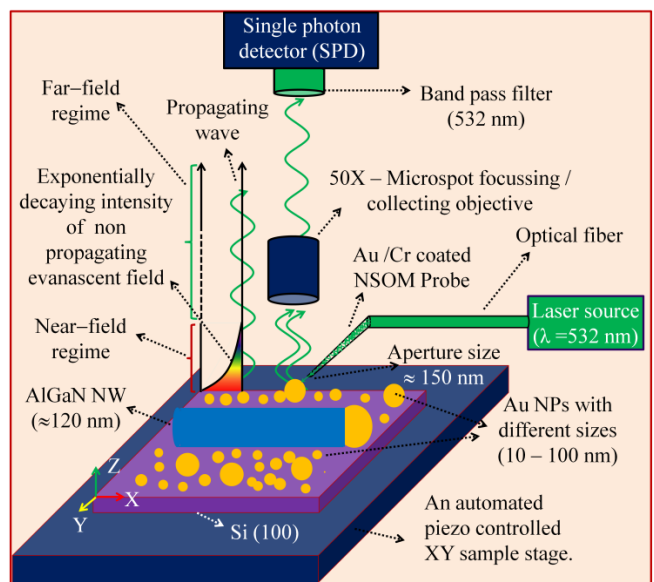

FIG. 2: The schematic experimental setup for NSOM imaging of AlGaN NW and Au NPs.

In order to perform an NSOM experiment, with near–field excitation and far–field collection configuration (FIG. 2), a point light source emanated through the probe at the near–field regime was scanned over the surface of the sample specimen with tuning fork feedback and the propagating optical signal emitted from the sample surface due to the dipole radiation was detected at the far–field. A band pass filter (532 nm) was employed to extract the excitation laser after the light–matter interaction and before reaching the light to the single photon detector (SPD) in the

far–field configuration. We used an optical fiber with a circular aperture and metal (Au/Cr) coated probe with a tip apex (aperture) diameter of 150 nm for near–field excitation of laser light. The optical fiber coated with Cr (buffer layer) and Au (over layer) to avoid leakage of optical power, was used to enhance the optical transmission and confine the light to the sample surface. The experimental setup (SPM with coupled Raman spectroscopy system) involved in the AFM/NSOM imaging system is shown in the figure 3.

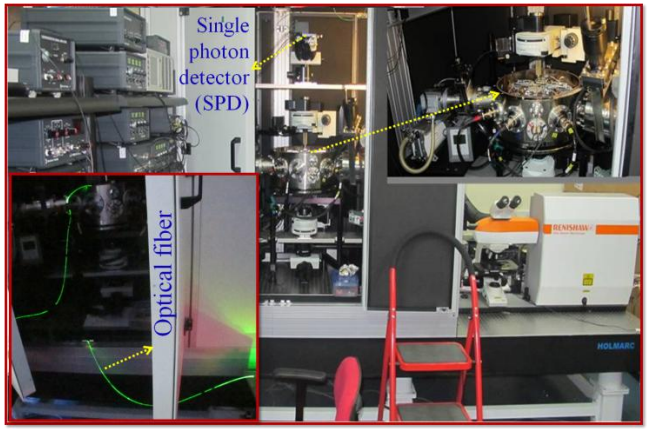

FIG. 3: The experimental set up; SPM (MultiView 4000; Nanonics, Israel) coupled Raman spectroscopy (Renishaw, UK) system, used for the AFM/NSOM imaging of the sample. The inset shows an optical fiber carrying laser light of 532 nm (2.33 eV) to NSOM probe with an aperture of 150 nm.

The same instrument with NSOM configuration is used to understand the near–field light–matter interactions of nanostructures with visible laser 532 nm (~2.33 eV). The scanning was performed either using the translational movement of NSOM probe (description for FIG. 2) or by motorized XY sample stage with very precise spatial resolution controlled by inbuilt sensors and piezo–drivers. A band pass filter (532 nm) was used to extract the scattered light, post light–matter interaction, before entering the SPD in the far field configuration. The same probe was used as an AFM tip for simultaneous scanning of the topography (description for FIG. 1(b)), along with the NSOM image of the sample with the tuning fork feedback mechanism.

## 3. Results and discussions

The morphological shape, size and distribution of mono–dispersed AlGaN NWs are shown in the AFM topographic image (FIG. 4(a)). The high resolution AFM image shows cylindrical shaped NWs with very smooth surface morphology along with Au NP catalyst at the tip (FIG. 4(b)). Uniformly sized and mono–dispersed NWs with average diameter of 120 nm were observed. The well separated Au NPs, which participated in the VLS growth process of the NWs, were uniformly sized (~150 nm) (FIG. 4(c)). The Au NPs with diameters around 50–100 nm, not participating in the growth process, were also found to be distributed uniformly over the substrate (FIG. 4(a)).

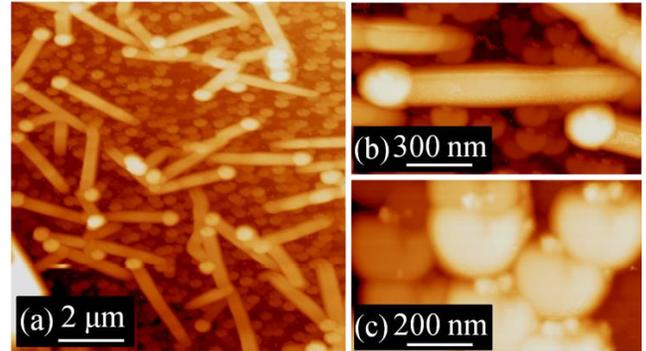

FIG. 4: AFM topographic images (a) mono–dispersed AlGaN nanowires and high resolution image of (b) single AlGaN nanowire (c) Au nanoparticles

Since the diameter of Au NPs (~100 nm) as well as AlGaN NW (~120 nm) is far below the diffraction limit for the excitation wavelength (532 nm), one needs to shorten the wavelength down to the sub–diffraction regime to obtain highly resolved optical images. Using metal coated NSOM probe, it is possible to produce evanescent waves with momentum higher than that of the original excitation wavelength $\lambda_0 = 2\pi/k_0(\omega)$ with wave vector of $k_0(\omega) = \omega/c$, where $c$ is the velocity of light. Therefore, the evanescent waves emanating from the NSOM probe aperture possess group of wave vectors higher than the original excitation laser as $k_{ev}(\omega) = \omega/v$, with different velocities ($v$) slower than the excitation wave velocity ($v<c$) [1-3]. Therefore, the NSOM measurements are advantageous in providing the super–resolution along with localization of intense electric fields. At the same time, it conserve the excitation energy and hence the frequency. Thus, it offers the possibility of optical as well as spectroscopic imaging in the sub–wavelength regime. So, from the scanned images using NSOM technique, we can understand even the intrinsic properties of a sample as revealed by its electronic or vibronic characteristics of the material.

The observed NSOM images (FIG. 5) of Au NPs shows a strong SPR related absorption in two dimension (2D) and 3D. The high resolution topographic AFM image of the Au NPs shows (2D and 3D images in FIGs. 5(a) and 5(b), respectively) smooth and spherical shape of the NP with a diameter of ~100 nm.

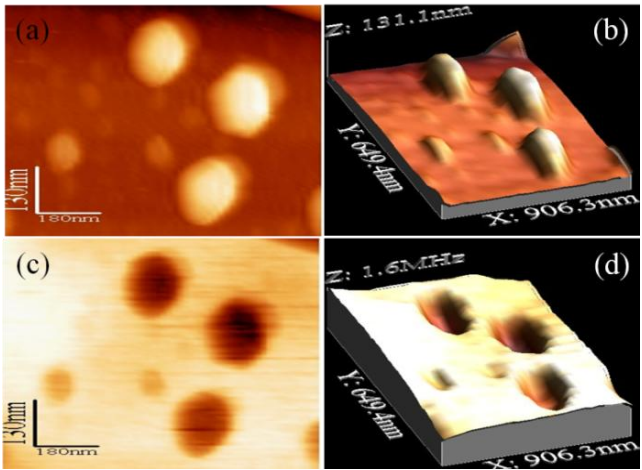

FIG. 5: AFM topographic images of Au nanoparticles in (a) 2D and (b) 3D. The NSOM images of Au nanoparticles in (a) 2D and (b) 3D.

The NSOM image of Au NPs shows (2D and 3D images in FIGs. 5(c) and 5(d), respectively) a strong absorption of electromagnetic waves. The significant absorption of light with wavelength 532 nm by Au NP is because of the fact that, the SPR peak value ~540 nm for Au NPs matches the excitation wavelength. At resonance, the incident electromagnetic waves coupled with collective oscillation of electrons, can produce SPPs which are perpendicular to the surface of the Au NP. The frequency dependent wave vector of SPP can be expressed in terms of frequency dependent dielectric constants of metal ($\varepsilon_m = \varepsilon_m' + i\varepsilon_m''$) and surrounding dielectric material ($\varepsilon_d = 1$, for air or vacuum), $k_{spp}(\omega) = \frac{\omega}{c}\sqrt{\frac{\varepsilon_d \cdot \varepsilon_m}{\varepsilon_d + \varepsilon_m}}$. Therefore, the effective wavelength of the SPP is $\lambda_{spp} = 2\pi/k_{spp}$ [1-3]. The SPPs of different wavelengths, lower than the excitation, can propagate through the surface of Au NPs up to a propagation length which depends on the complex dielectric constants of the metal and dielectric medium [1-3]. Once the SPP propagates through the surface of Au NP and crosses the metallic region, then the electromagnetic wave may decouple from the SPP and it can be converted to a propagating wave. The intensity of the absorption is influenced by the frequency dependent polarizability of the Au NPs and it can vary with respect to the size of the Au NPs. Thus, because of the variation of different sizes of the Au NPs, it is possible to observe them with relatively different absorption intensities (FIGs. 5(c) and 5(d)). Apart from the formation of SPP, some portion of the absorbed excitation laser may also participate in lattice phonon generations leading to heating as well as inter–band transitions of Au NPs [7].

The NSOM images obtained as a result of near–field light–matter interaction is shown for AlGaN single NW along with Au NPs of various sizes (FIG. 6). The high resolution topographic AFM image of the single NW shows (2D and 3D images in FIGs. 6(c) and 6(d), respectively) smooth and cylindrical shape, as observed in the FESEM images. The NSOM images of AlGaN single NW as well as catalyst Au NPs are also observed (2D and 3D images in FIGs. 6(c) and 6(d), respectively). The reported room temperature band gap of our AlGaN NWs is 3.55 eV [5], which is higher than the excitation energy of 2.33 eV. Therefore, a complete transmission of light through the AlGaN NW is expected. Surprisingly, we observed a prominent absorption of light along the AlGaN NW, as shown in the NSOM images (2D and 3D images in FIGs. 6(c) and 6(d), respectively).

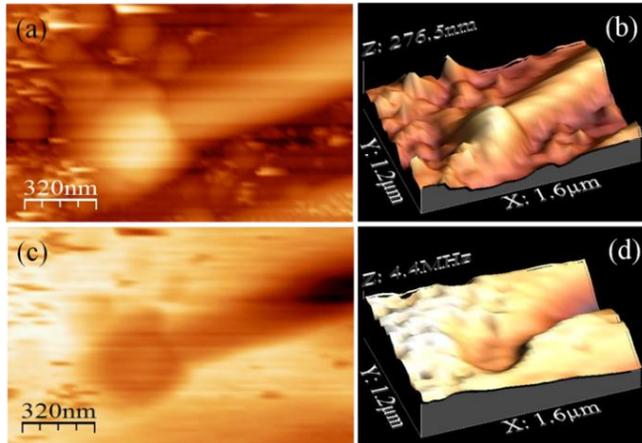

FIG. 6: AFM topographic images of mono–dispersed AlGaN nanowire in (a) 2D and (b) 3D. The NSOM images of AlGaN nanowire in (a) 2D and (b) 3D.

This absorption of light by the semiconductor AlGaN NW is observed because of the presence of native defects originating due to the unavoidable incorporation of C and O in the material, which may create energy levels below 2.33 eV.

In conclusion, we envisage the use of near field scanning optical microscopy (NSOM) technique for direct understanding of light–matter interaction of metallic as well as semiconductor nanostructures of sub–wavelength limited dimension in the near–field regime. The NSOM images of metallic Au nanoparticles with diameter ~100 nm shows a strong surface plasmon resonance related absorption of excitation laser with an energy of 2.33 eV (532 nm) due to the formation of surface plasmon polaritons as well as the localized surface plasmon resonance near to the surface of the Au nanoparticles. The isolated single semiconductor AlGaN nanowire with a diameter ~120 nm shows a strong absorption of visible light due to the electronic transitions originated from the native defect related energy levels.

## Acknowlegements

We thank, S. Dhara of SND, IGCAR for his valuable guidance and helpful discussions.